# Slow White Dwarf Mergers as a New Galactic Source of Trans-Iron Elements

*Umberto* Battino[1,2,*], *Claudia* Lederer-Woods[2], *Claudia* Travaglio[3], *Friedrich Konrad* Röpke[4], and *Brad* Gibson[1]

[1]E.A. Milne Centre for Astrophysics, University of Hull, Hull, HU6 7RX, UK
[2]School of Physics and Astronomy, University of Edinburgh, Edinburgh, EH9 3FD, UK
[3]INFN - Istituto Nazionale Fisica Nucleare, 10125 Turin, Italy
[4]Heidelberg Institute for Theoretical Studies, 69118 Heidelberg, Germany

**Abstract.** The astrophysical origins of the heaviest stable elements that we observe today in the Solar System are still not fully understood. Recent studies have demonstrated that H-accreting white dwarfs (WDs) in a binary system exploding as type Ia supernovae could be an efficient *p*-process source beyond iron. However, both observational evidence and stellar models challenge the required frequency of these events. In this work, we calculate the evolution and nucleosynthesis in slowly merging carbon-oxygen WDs. As our models approach the Chandrasekhar mass during the merger phase, the $^{22}$Ne($\alpha$,n)$^{25}$Mg neutron source reaction is activated in the external layers of the primary WD, where the carbon-rich material accreted from the secondary WD is burned via the $^{12}$C+$^{12}$C reaction, which provides the necessary $\alpha$-particles via the $^{12}$C($^{12}$C,$\alpha$)$^{20}$Ne channel. The resulting neutron capture abundance distribution closely resembles a weak *s*-process one and peaks at Zr, which is overproduced by a factor of 30 compared to solar. The mass of the most external layers enriched in first-peak *s*-process elements crucially depends on the $^{12}$C+$^{12}$C reaction rate, ranging between 0.05 M$_\odot$ and ∼0.1 M$_\odot$. These results indicate that slow white dwarf mergers can efficiently produce the lightest *p*-process isotopes (such as $^{74}$Se, $^{78}$Kr, $^{84}$Sr, $^{92}$Mo and $^{94}$Mo) via $\gamma$-induced reactions if they explode via a delayed detonation mechanism, or eject the unburned external layers highly enriched in first peak s-process elements in the case of a pure deflagration. In both cases, we propose for the first time that slow WD mergers in binary systems may be a new relevant source for elements heavier than iron.

## 1 Introduction

Type Ia supernovae (SNIa), the violent death of white-dwarfs in a binary system, play a central role in the chemical evolution of galaxies. Despite their importance, the nature of SNIa progenitors is still a matter of debate, with two main scenarios proposed. In the single-degenerate scenario (see [1]), a WD accretes mass from the envelope of a less evolved stellar companion. When the object approaches the critical Chandrasekhar-mass (the maximum mass that can be sustained by electron degeneracy, around 1.39 M$_\odot$ for CO WD), C-burning ignites explosively causing a SNIa. An exception can occur when the accreted matter is mainly He, which would trigger a detonation before the Chandrasekhar mass is reached (sub-Chandrasekhar SNIa, [2]). In the double-degenerate scenario (see [3]), the companion is another WD. Over time, the separation of the pair decreases until they merge, either via a violent merger event or the formation of an accretion disk around the primary WD if the mass ratio of the components is larger or lower than ∼0.8, respectively.

---

*e-mail: μU.Battino@hull.ac.uk

The relevance of SNIa in galactic chemical evolution (GCE) depends on the fraction of SNIa made via each of the aforementioned channels. [4] showed how SNIa originating from the single-degenerate scenario can be a very important source of a class of trans-iron proton-rich isotopes known as *p*-nuclei. The main assumption of [4], recently confirmed by [5], was the presence of a trans-iron element distribution up to Pb (called "seed distribution") synthesized on the WD surface during the accretion of H-rich material from the companion. The formation of the surface seed distribution is crucial, as γ-induced reactions acting on it during the explosion can result in the efficient synthesis of the *p*-nuclei. Moreover, [4] found that if about 50–70% of all SNIa are made via the single-degenerate scenario, they can be responsible for at least 50% of the *p*-nuclei abundances in the Solar System.

On the other hand, this is not what was obtained by [6] who, based on the spectra from a large sample of early-type galaxies, found that the single-degenerate SNIa should be less than 6% of the total SNIa population. Furthermore, [7] computed the nucleosynthetic ejecta of sub-Chandrasekhar SNIa from the helium detonation scenario, re-determining the contributions of both near-Chandrasekhar and sub-Chandrasekhar events to the total SNIa rate by reproducing the observed [Ti/Fe], [Mn/Fe] and [V/Fe] in the Galaxy. They found that at least 30% of all SNIa should come from near-Chandrasekhar-mass explosions. This result has been recently qualitatively confirmed by [8], who showed how there may be a very small contribution needed from near-Chandrasekhar-mass explosions only to explain the solar [Mn/Fe], and at the same time how this conclusion, however, depends on the massive star yields adopted in the calculations. Therefore, if the relative contribution of the single-degenerate channel to the total SNIa rate is observationally constrained to be below ∼6%, then ∼24% of near Chandrasekhar-mass SNIa is coming from an alternative channel. Can near-Chandrasekhar-mass SNIa originate from double degenerate systems? This is still an open question, with two primary shortcomings identified. First, during the disk formation, carbon could be ignited near the accreting star's surface and the burning transferred very rapidly towards the inner regions, thus producing an ONeMg WD, and not a thermonuclear supernova. This has been excluded by [9], who found that C-burning indeed occurs, but it is soon extinguished, unless more than ∼0.01 $M_\odot$ of He is present (see [10]). The second shortcoming consists of the fact that the mass transfer from the disc to the primary WD may occur at a high rate (∼$10^{-5}$ $M_\odot$ yr$^{-1}$), causing an off-centre explosive ignition of C-burning, well before the WD attains the Chandrasekhar mass. However, [11] showed that the effects of rotation, naturally arising in merging WDs, make the accretion rate *self-regulated* and as low as a few $10^{-7}$ $M_\odot$yr$^1$.

In this work, we assumed that Chandrasekhar-mass WDs can actually be formed via the double-degenerate channel through a slow merger process via the accretion-disc, which we simulate using the stellar evolution code MESA (see [12]). We want to verify if a trans-iron seed distribution can be formed on the surface of the pre-supernova structure, which may result in a novel source of trans-iron elements in the Universe.

## 2 Computational methods and numerical results

The stellar models presented in this section are computed using the 1D stellar code MESA (revision 10108). The CO WD model 0.927-from-6.0-z2m2.mod, included in the MESA data folder, was cooled down for $10^5$yr before being used as a starting model for our calculations.

The chemical composition in mass fraction of the accreted material, added at a constant accretion rate $\dot{M}$ =2.4 $10^{-6}$ M$_\odot$yr$^{-1}$, consisted of 0.48809 of $^{12}$C, 0.48809 of $^{16}$O, 0.002 of $^{20}$Ne, 0.021 of $^{22}$Ne, and the remaining 0.00082 in iron-group isotopes between $^{50}$Cr and $^{64}$Ni (with X($^{56}$Fe)=0.0007). The nuclear reaction network used in our simulations is co-burn-extras.net, which mainly includes the isotopes needed for carbon burning up to $^{28}$Si coupled by more than 50 reactions. The nuclear reaction rates recommended in the JINA REACLIB library were adopted (see [13]), except for the $^{12}$C+$^{12}$C reaction rate from [14], the $^{12}$C($\alpha,\gamma$)$^{16}$O from [15], and the $^{22}$Ne($\alpha$,n)$^{25}$Mg and $^{22}$Ne($\alpha,\gamma$)$^{26}$Mg from [16]. We extended the nuclear reaction network up to $^{141}$Ba when the accreting WD approaches the Chandrasekhar mass and the temperature at the base of the accreted material exceeds $10^9$K, in order to follow the neutron capture nucleosynthesis during the surface C-burning.

In Figure 1 the temperature and density profiles of our accreting WD are presented at different moments during the accretion towards the Chandrasekhar mass, closely resembling the ones obtained by [17]. When the WD mass reaches ∼1.37 M$_\odot$, the temperature at the base of the accreted material exceeds $10^9$K as a consequence of surface C-burning ignition. α particles are released via the $^{12}$C($^{12}$C, α)$^{20}$Ne reaction, which are then captured by $^{22}$Ne and converted into neutrons via the $^{22}$Ne($\alpha$,n)$^{25}$Mg reaction.

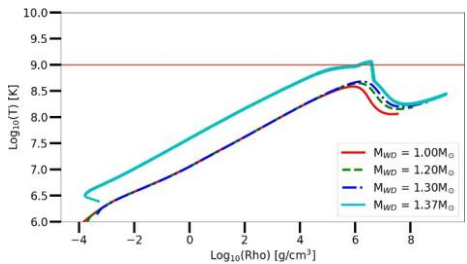

**Figure 1.** Temperature and density profiles of our accreting WD at different moments during the accretion towards the Chandrasekhar mass.

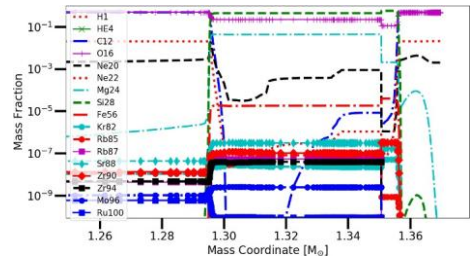

**Figure 2.** The neutron-capture isotopes produced on the top of the WD at the end of surface C-burning when adopting a $^{12}$C+$^{12}$C reaction rate increased by a factor of two.

The neutron-capture seed products on the top of the WD at the end of surface C-burning are shown in Figure 2. Given the key role and still high uncertainty of the $^{12}$C+$^{12}$C reaction rate, we performed two simulations, one adopting our standard rate and a second one with the reaction rate increased by a factor of two. In both cases, trans-iron elements are synthesized on the WD surface, forming essentially a weak *s*-process distribution peaked at Zr, which is overproduced by a factor of 30 compared to solar, and up to Mo. Moreover, depending on the $^{12}$C+$^{12}$C reaction rate, the thickness in mass of the surface material enriched in heavy elements is 0.05 and 0.07 M$_\odot$ when adopting the standard and the increased $^{12}$C+$^{12}$C reaction rate, respectively. This size is of the same order of what was found by [4] to be a prerequisite of an exploding near-Chandrasekhar WD in order to trigger an efficient *p*-process, and could include the production of the historically puzzling cases of Mo *p*-isotopes.

## 3 Conclusions

In this work, we calculated the evolution and nucleosynthesis in slowly merging CO WDs using the stellar code MESA. As our models approach the Chandrasekhar mass during the merger phase, the $^{22}$Ne(,n)$^{25}$Mg neutron source reaction is activated in the external layers of the primary WD. The resulting neutron capture seed distribution closely resembles a weak *s*-process one and peaks at Zr, which is overproduced by a factor of 30 compared to solar. The mass of the most external layers enriched in first-peak *s*-process elements ranges between 0.05 M$_\odot$ and ∼0.1 M$_\odot$, which would allow an efficient *p*-process during the subsequent explosion. Hence slow WD mergers could really be a novel trans-iron elements source, potentially being able to efficiently produce the lightest *p*-process isotopes (such as $^{74}$Se, $^{78}$Kr, $^{84}$Sr, $^{92}$Mo and $^{94}$Mo) via γ-induced reactions if they explode via a delayed detonation mechanism, or eject the unburned external layers highly enriched in first peak *s*-process elements in the case of a pure deflagration. We then propose for the first time that slow WD mergers may be a new relevant astrophysics source for elements heavier than iron, pending a closer investigation by simulating the explosive nucleosynthesis originating from these progenitors. It must be stressed that only one solar-scaled metallicity composition of the accreted material has been studied so far. We will extend our investigation to different metallicities, as well as different C/O ratios and $^{22}$Ne abundance in both the primary and secondary WD.

This work received support from the Science and Technology Facilities Council UK (ST/M006085/1 & ST/R000840/1), and the European Research Council ERC-2015-STG Nr. 677497. This article is based upon work from the ChETEC COST Action (CA16117). UB & BKG acknowledges support by ChETEC-INFRA (EU project no. 101008324).